\documentclass[twocolumn,aps,amsmath,amssymb,superscriptaddress,prx]{revtex4-1}   
\usepackage[latin2]{inputenc}
\usepackage{amsmath,amssymb}
\usepackage{amsfonts}
\usepackage{bm}
\usepackage{dcolumn}
\usepackage{setspace}
\usepackage{graphicx}
\usepackage{color}
\usepackage{multirow}
\usepackage{verbatim}
\usepackage{epstopdf}

\def\bar{\begin{array}}
\def\ear{\end{array}}

\def\f{\frac}

\def\nn{\nonumber}

\def\m{\omega^*}

\def\k{\mathbf{k}}

\def\p{\partial}
\def\Re{\mathrm{Re}}

\def\a{\alpha}

\def\l{\lambda}
\def\k{\kappa}
\def\m{\mu}
\def\n{\nu}
\def\r{\rho}

\begin{document}

\title{Quantum covariant derivative}

\author{Ryan Requist} 
\affiliation{Fritz Haber Center for Molecular Dynamics, Institute of Chemistry, The Hebrew University of Jerusalem, Jerusalem 91904 Israel}

\date{\today}

\begin{abstract}
The covariant derivative capable of differentiating and parallel transporting tangent vectors and other geometric objects induced by a parameter-dependent quantum state is introduced.  It is proved to be covariant under gauge and coordinate transformations and compatible with the quantum geometric tensor.  The quantum covariant derivative is used to derive a gauge- and coordinate-invariant adiabatic perturbation theory, providing an efficient tool for calculations of nonlinear adiabatic response properties.
\end{abstract}

\maketitle

\section{Introduction}\label{sec1}

Quantum geometry, particularly the Riemannian geometry of a parameter-dependent quantum state, is surging into mainstream condensed matter physics.  In fact, the discovery of a gauge-invariant Riemannian metric \cite{provost1980}, called the quantum metric, came after decades of effort \cite{bohr1953,inglis1954,villars1958,thouless1962,rowe1974,reinhard1979}
in nuclear physics aimed at deriving a reduced description of collective nuclear rotation that accounts for the dynamics of the constituent nucleons.

There are clear analogies to molecular rotation and its coupling to vibrations and electrons. Physical effects of the quantum metric were studied in generic fast-slow systems modeled on the Born-Oppenheimer approximation in molecular physics \cite{berry1989,berry1990,aharonov1992,berry1993}, where the parameters are atomic coordinates and the quantum metric produces an effective scalar potential in the Schr\"odinger equation for the nuclei.  The quantum metric is thus connected with effective electric fields \cite{berry1989}, while the Berry curvature is associated with effective magnetic fields \cite{mead1979,berry1984}.  The quantum metric and Berry curvature derive from a single complex-valued tensor, called the quantum geometric tensor.  It is not necessary to rely on the Born-Oppenheimer approximation to define a quantum geometric tensor: a nonadiabatic quantum geometric tensor has been defined \cite{requist2016a} using the exact factorization method \cite{gidopoulos2014,abedi2010,hunter1975}.  This quantum metric is indispensible in the nonadiabatic generalization of density functional theory \cite{requist2016b,li2018}, where it is the source of the phonon-induced kink in photoemission spectroscopy \cite{requist2019,boeri2022}.  The effective scalar potential, which in the Born-Oppenheimer approximation is divergent at conical intersections \cite{berry1989}, has been regularized in the context of molecular dynamics \cite{rawlinson2020}, and an analogous quantum metric has been studied in purely electronic systems \cite{schild2017,kocak2021}.

Band structure theory is another area where the quantum metric has appeared. Quantities such as Wannier function width \cite{marzari1997}, field induced position shifts \cite{gao2014}, orbital magnetic susceptibility \cite{gao2015}, and angular momentum-dependent excitonic shifts \cite{srivastava2015}, the superfluid weight of flat bands \cite{liang2017}, and the injection current in topological semimetals \cite{ahn2020} have been shown to depend on the quantum metric.  Quantum geometry can be extended beyond mean-field theory using the concepts of natural orbital geometric phase and natural orbital band structure \cite{requist2012,requist2018,requist2021}.  The quantum geometric tensor has also been shown to be a useful marker of quantum phase transitions in spin chains \cite{carollo2005,zhu2006,campos-venuti2007,carollo2020}.  

In this article, we introduce the quantum covariant derivative, a geometric concept that unifies the gauge-covariant derivative and the coordinate-covariant derivative and provides a foundation for further developments in quantum geometry.  The purpose of the quantum covariant derivative, denoted $\hat{\nabla}$, is to differentiate the geometric objects that arise in quantum mechanics and define the parallel transport of tangent vectors to quantum state space.  We prove that it satisfies a condition analogous to that satisfied by the covariant derivative in Riemannian geometry, namely $\hat{\nabla}$ is compatible with the quantum geometric tensor in the sense that parallel transport by $\hat{\nabla}$ preserves the Hermitian inner product. 

As a first application, we use $\hat{\nabla}$ to derive a novel adiabatic perturbation theory that is gauge and coordinate invariant.  This result is of fundamental importance for the systematic calculation of nonlinear response properties and extends the geometric and topological formulations of adiabatic linear response, e.g.~adiabatic charge transport \cite{laughlin1981,thouless1982,avron1983,thouless1983,niu1984,resta1992,king-smith1993,ortiz1994}.

\section{Covariant differentiation}

We start by considering a quantum state $|\psi\rangle=|\psi(x)\rangle$ depending parametrically on parameters $\{x^{\mu}\}$, $\mu=1,\ldots,n$, that provide local coordinates in a manifold $M$.  This covers many situations.  Cases of special interest occur when (i) $|\psi\rangle$ is a cell-periodic Bloch function $| u_{n\mathbf{k}}\rangle$ and $x^{\mu}$ are the components of the wavevector $\mathbf{k}$, (ii) $|\psi\rangle$ is an electronic state and $x^{\mu}$ are generalized nuclear coordinates, and (iii) $|\psi\rangle$ is a many-nucleon state and $x^{\mu}$ are parameters that define the shape of the nucleus.  

A parameter-dependent state such as $|\psi(x)\rangle$ has gauge freedom, i.e.~the state $|\tilde{\psi}(x)\rangle = e^{if(x)} |\psi(x)\rangle$ provides an equally valid description of the quantum system.  The gauge-covariant derivative 
\begin{align}
D_{\mu} = \p_{\mu} + i A_{\mu} {,}
\label{eq:covariant:derivative}
\end{align}
where $\p_{\mu}=\p/\p x^{\mu}$ and $A_{\mu} = i \langle \psi | \p_{\mu} \psi\rangle$, gives us a way to differentiate a state $|\psi\rangle$ such that $D_{\mu} |\psi\rangle$ and $|\psi\rangle$ are {\it gauge covariant}, i.e.~they transform in the same way under the gauge transformation
\begin{align}
|\psi\rangle &\rightarrow |\tilde{\psi}\rangle = e^{i f(x)} |\psi\rangle {;} \nn \\
D_{\mu} |\psi\rangle &\rightarrow D_{\mu} |\tilde{\psi}\rangle = e^{i f(x)} D_{\mu}|\psi\rangle {.}
\label{eq:gauge:covariance}
\end{align}

The gauge-covariant derivative $D_{\mu}$ is one of several kinds of covariant derivatives.  Covariant derivatives differentiate geometric objects in an intrinsic way, i.e.~in a way that is independent of the arbitrary choices that are made in actual computations.  The prime example is the covariant derivative $\nabla$ in Riemannian geometry, which is independent of the choice of local coordinates.  Because quantum states have gauge freedom, what distinguishes the quantum covariant derivative $\hat{\nabla}$ from the covariant derivative $\nabla$ is the need to require $\hat{\nabla}$ to be simultaneously gauge and coordinate covariant. 

For a quantum state depending on $\{x^{\mu}\}$, the kets
\begin{align}
| D_{\m}\psi\rangle = \bigg| \f{\p\psi}{\p x^{\m}} \bigg> - |\psi\rangle \bigg<\psi \bigg| \f{\p\psi}{\p x^{\m}} \bigg> {,} \quad \m=1,\ldots,n {,}
\end{align}
assumed to be linearly independent throughout a region of interest in $M$, provide a gauge-covariant basis for a complex vector space $E$ at each point in $M$.  A general vector $| v\rangle = | D_{\nu}\psi\rangle v^{\nu}$ (we use the summation convention) in $E$ will be called a {\it tangent ket} to distinguish it from a tangent vector to $M$.  The projection of $| D_{\mu}\psi\rangle$ onto $|\psi\rangle$ is zero by virtue of the normalization of $|\psi\rangle$, which we assume, and therefore $|\psi\rangle$ plays a role similar to the unit normal vector to a surface embedded in Euclidean space in the context of Riemannian geometry.

As a first step in defining the quantum covariant derivative, we decompose the $D_{\m}$-derivative of $| D_{\n} \psi \rangle$ into normal and tangential components according to
\begin{align}
D_{\m} | D_{\n} \psi \rangle &= |\psi\rangle \langle \psi | D_{\m} D_{\n} \psi \rangle + | D_{\l} \psi\rangle \Upsilon^{\l}_{\m\n} {,}
\label{eq:decomposition}
\end{align}
where $\Upsilon^{\l}_{\m\n}$ is a rank-3 quantity, which we call the quantum Christoffel symbol of the second kind.  Since $|\psi\rangle$, $| D_{\n}\psi\rangle$, and $| D_{\m} D_{\n} \psi\rangle$ all transform gauge covariantly, in the way displayed in Eq.~(\ref{eq:gauge:covariance}), $\Upsilon^{\l}_{\m\n}$ is gauge invariant.  

Take $\mathcal{C}$ to be a path in $M$ parametrized by $t$, and let $| v\rangle =| v(t)\rangle$ be a tangent ket vector field defined on the image of $\mathcal{C}$.  The derivative 
\begin{align}
\bigg| \f{dv}{dt} \bigg> = | D_{\n} \psi\rangle \f{dv^{\n}}{dt} + | D_{\m} D_{\n}\psi\rangle \f{dx^{\m}}{dt} v^{\n} - | D_{\n} \psi\rangle  i A_{\m} \f{dx^{\m}}{dt} v^{\n}
\end{align}
at a point along $\mathcal{C}$ is generally not an element of the vector space $E$ at that point.  Projecting $| dv/dt \rangle$ into $E$ yields the coordinate-invariant derivative 
\begin{align}
\f{\mathcal{D} | v \rangle}{dt}  = | D_{\l}\psi\rangle \bigg( \f{dv^{\l}}{dt} + \omega_{\m\n}^{\l} \f{dx^{\m}}{dt} v^{\n} \bigg) {}
\end{align}
with the coefficients $\omega_{\m\n}^{\l} = \Upsilon_{\m\n}^{\l} - i A_{\m} \delta^{\l}_{\n}$.  If a gauge transformation $|\psi\rangle\rightarrow e^{if(x)} |\psi\rangle$ is applied along $\mathcal{C}$, the coordinate-covariant derivative transforms as 
\begin{align}
\f{\mathcal{D} | v \rangle}{dt} \rightarrow \f{\mathcal{D} | \tilde{v} \rangle}{dt} = e^{if} \f{\mathcal{D} | v \rangle}{dt} + i \f{\p f}{\p x_{\m}} \f{dx^{\m}}{dt} e^{if} | v\rangle {.}
\end{align}
Compensating for the $df/dt$ term, we define the gauge- and coordinate-covariant derivative
\begin{align}
\f{\hat{\nabla} | v \rangle}{dt} &= \f{\mathcal{D} | v \rangle}{dt} + i A_{\m} \f{dx^{\m}}{dt}| v\rangle \nn \\
&= | D_{\l}\psi\rangle \bigg( \f{dv^{\l}}{dt} + \Upsilon_{\m\n}^{\l} \f{dx^{\m}}{dt} v^{\n} \bigg) {.}
\label{eq:nabla:def}
\end{align}
The quantum covariant derivative of $| v\rangle$ with respect to $\mathbf{X}$, an arbitrary vector tangent to $M$, is
\begin{align}
\hat{\nabla}_{\bf X} | v\rangle = | D_{\l}\psi\rangle \bigg( \f{\partial v^{\l}}{\partial x^{\m}} + \Upsilon^{\l}_{\m\n} v^{\n} \bigg) X^{\m} {.}
\label{eq:nabla}
\end{align}
The quantum Christoffel symbol $\Upsilon^{\l}_{\m\n}$ specifies the quantum covariant derivative $\hat{\nabla}$ in a coordinate frame for $M$.  Equation (\ref{eq:nabla}) is our main result.

Except for the crucial distinction that $\Upsilon^{\l}_{\m\n}$ is generally complex, Eq.~(\ref{eq:nabla}) is analogous to the Riemannian covariant derivative $\nabla_{\mathbf{X}} \mathbf{v}$ of a vector field $\mathbf{v}=\mathbf{e}_{i}v^{i}$ with respect to $\mathbf{X}=\mathbf{e}_{i}X^{i}$, i.e. 
\begin{align}
\nabla_{\mathbf{X}} \mathbf{v} = \mathbf{e}_{i} \bigg( \f{\p v^{i}}{\p x^j} + \Gamma^{i}_{jk} v^{k} \bigg) X^{j} {,}
\label{eq:covariant:derivative:classical}
\end{align}
where $x^{i}$ are local coordinates, $\mathbf{e}_{i}$ are basis vectors, and $\Gamma^{i}_{jk}$ is the Christoffel symbol, which is computable in terms of the Riemannian metric $g_{ij}$.  The covariant derivative in Eq.~(\ref{eq:covariant:derivative:classical}) is the unique symmetric covariant derivative that is compatible with the metric $g_{ij}$ in the sense that the inner product $g(\mathbf{u},\mathbf{v})=u^i g_{ij}v^j$ of any two vectors $\mathbf{u}$ and $\mathbf{v}$ is preserved when those vectors are parallel transported \cite{ricci1900,wald1984,frankel1997}.  

We now prove that (i) $\hat{\nabla}$ is a symmetric covariant derivative, i.e.~$\Upsilon^{\l}_{\m\n}=\Upsilon^{\l}_{\n\m}$, and (ii) $\hat{\nabla}$ is compatible with the quantum geometric tensor.  Hence, $\hat{\nabla}$ is the quantum mechanical generalization of the covariant derivative in Riemannian geometry.

The space of quantum states parametrized by $x^{\mu}$ has Riemannian structure but also additional geometric structure defined by the quantum geometric tensor \cite{provost1980,berry1989}
\begin{align}
h_{\m\n} &= \bigg< \f{\p\psi}{\p x^{\m}} \bigg| \Big(1-\big|\psi\big> \big<\psi\big|\Big) \bigg| \f{\p\psi}{\p x^{\n}} \bigg> \nn \\
&= \langle D_{\m} \psi | D_{\n} \psi \rangle {.} \label{eq:h}
\end{align}
The real part of $h_{\m\n}$ defines the quantum metric $g_{\m\n}$.  The imaginary part is minus one-half the Berry curvature $B_{\m\n}=\p_{\m} A_{\n}-\p_{\n} A_{\m}$.   The quantum geometric tensor is a Hermitian metric that assigns, to each point in $M$, a complex inner product,
\begin{align}
h(| u\rangle,| v\rangle) = \overline{u}^{\alpha} h_{\alpha\beta} v^{\beta} {,}
\label{eq:innerproduct}
\end{align}
on the complex vector space $E$ spanned by the $| D_{\m}\psi\rangle$; an overline denotes complex conjugation.  For $\hat{\nabla}$ to be compatible with $h_{\m\n}$ means that 
\begin{align}
\f{dh(| u\rangle,| v\rangle)}{dt} = h\bigg(| u\rangle,\f{\hat{\nabla} | v\rangle}{dt}\bigg) + h\bigg(\f{\hat{\nabla} | u\rangle}{dt},| v\rangle\bigg) {,}
\label{eq:compatibility:1}
\end{align}
where $| u\rangle$ and $| v\rangle$ are any smooth tangent kets defined along a path $\mathcal{C}$ in $M$ \cite{tu2017}. 

We first show that $\hat{\nabla}$ is symmetric.  It is convenient to define the quantum Christoffel symbol of the first kind,
\begin{align}
\Upsilon_{\l\m\n} &\equiv \langle D_{\l} \psi | D_{\m} D_{\n} \psi \rangle \nn \\
&= h_{\l\r} \Upsilon^{\r}_{\m\n} {,}
\label{eq:Upsilon:relation}
\end{align}
where the second line follows from Eqs.~(\ref{eq:decomposition}) and (\ref{eq:h}).  Hence, the quantum geometric tensor $h_{\l\r}$ lowers the upper index of $\Upsilon^{\r}_{\m\n}$ just as the Riemannian metric $g_{\l\r}$ lowers the upper index of $\Gamma^{\r}_{\m\n}$.  
The symmetry of $\Upsilon^{\l}_{\m\n}$ with respect to interchange of $\m$ and $\n$ follows immediately from the corresponding symmetry of $\Upsilon_{\l\m\n}$, which has already been proven \cite{requist2022a}.

Unlike the matrix $(g_{\l\r})$, which in Riemannian geometry is invertible by assumption, the matrix $(h_{\l\r})$ is not guaranteed to be invertible.  However, in this section we shall assume that $h_{\m\n}$ is invertible in a region of interest in $M$, so its inverse, $h^{\l\r}$, can be used to raise the first index of $\Upsilon_{\r\m\n}$, i.e.~$\Upsilon^{\l}_{\m\n} = h^{\l\r} \Upsilon_{\r\m\n}$.  In this way, the symbol $\Upsilon_{\l\m\n}$ leads to the definition of the quantum covariant derivative in Eq.~(\ref{eq:nabla}), as previously pointed out \cite{requist2022a}.  

To prove compatibility, use the product rule on the left hand side of Eq.~(\ref{eq:compatibility:1}) and invoke the arbitrariness of the kets $| u\rangle$ and $| v\rangle$ to find
\begin{align}
\p_{\m} h_{\l\n} &= h_{\l\r} \Upsilon^{\r}_{\m\n} + \overline{h}_{\n\r} \overline{\Upsilon}^{\r}_{\m\l} \nn \\
&= \Upsilon_{\l\m\n} + \overline{\Upsilon}_{\n\m\l} {.} \label{eq:identity}
\end{align}
Adding and subtracting this identity with different permutations of the indices, we obtain
\begin{align}
\f{\p h_{\l\n}}{\p x^{\m}} + \f{\p h_{\m\l}}{\p x^{\n}} - \f{\p h_{\n\m}}{\p x^{\l}}  = 2 \Re \Upsilon_{\l\m\n} &+ i 2 \mathrm{Im} (\Upsilon_{\m\l\n}-\Upsilon_{\n\l\m}) {,}\nn \\
\end{align}
which yields an identity
\begin{align}
\Re \Upsilon_{\l\m\n} = \f{1}{2} \bigg( \f{\p g_{\l\m}}{\p x^{\n}} + \f{\p g_{\n\l}}{\p x^{\m}} - \f{\p g_{\m\n}}{\p x^{\l}}\bigg)  \label{eq:identity:real}
\end{align}
that is satisfied by the $\Upsilon_{\l\m\n}$ defined in Eq.~(\ref{eq:Upsilon:relation}).  From the imaginary part of Eq.~(\ref{eq:identity}), we obtain the identity 
\begin{align}
\mathrm{Im} \Upsilon_{\l\m\n} - \mathrm{Im} \Upsilon_{\n\m\l} = -\f{1}{2} \f{\p B_{\l\n}}{\p x^{\m}} {,} \label{eq:identity:C:imag}
\end{align}
which is also satisfied by $\Upsilon_{\l\m\n}$ \cite{requist2022a}.  Since the right hand side of Eq.~(\ref{eq:identity:real}) is the standard formula for the classical Christoffel symbol of the first kind, $\Gamma_{\l\m\n}$, we can write $\Upsilon_{\l\m\n} = \Gamma_{\l\m\n} + i C_{\l\m\n}$, where $C_{\l\m\n}=\mathrm{Im} \Upsilon_{\l\m\n}$ is a quantity whose physical significance remains to be explored.  It appears in the equation of motion for $g_{\mu\nu}$ in the context of the coupled dynamics of electrons and nuclei \cite{requist2022a}.  

Unlike the covariant derivative $\nabla$, which is the unique symmetric covariant derivative compatible with $g_{\m\n}$, the quantum covariant derivative $\hat{\nabla}$ is not the unique symmetric connection compatible with $h_{\m\n}$.  The alternative symmetric symbol $\Lambda^{\l}_{\m\n} = h^{\l\r}(\Upsilon_{\r\m\n} + i F_{\r\m\n})$ defines a covariant derivative that is compatible with $h_{\m\n}$ if the rank-3 quantity $F_{\r\m\n}$ is fully symmetric.  Substituting $\Lambda_{\l\m\n}=h_{\l\r}\Lambda^{\r}_{\m\n}$ in Eq.~(\ref{eq:identity:C:imag}) shows that $F_{\l\m\n}$ must be symmetric in $\l\n$, which together with the assumed symmetry in $\m\n$, implies $F_{\l\m\n}$ must be a fully symmetric quantity.  Nevertheless, Eq.~(\ref{eq:Upsilon:relation}) defines a particular symbol $\Upsilon^{\l}_{\m\n}$ that is compatible with $h_{\m\n}$. 

Parallel transport is an important application of the covariant derivative.  The quantum mechanical geometric phase $\gamma = \oint A_{\mu} dx^{\mu}$ \cite{longuet-higgins1958,mead1979,mead1980c,berry1984,wilczek1984,aharonov1987,samuel1988,min2014,requist2017} arises from parallel transporting a quantum state \cite{simon1983}.  Similarly, the quantum covariant derivative can be used to parallel transport a tangent ket $| v\rangle$ via the rule $\hat{\nabla} | v\rangle = 0$.  The parallel transport equations for a three-state quantum system are derived in App.~A.  After parallel transport around a closed path in $M$ parametrized by $s$, the final tangent ket will differ from the initial tangent ket by the path-ordered exponential
\begin{align}
G = P\mathrm{exp}\bigg[ -\int_0^1 \mathbb{A}_{\mu}(s) (dx^{\mu}/ds) ds \bigg] {,}
\label{eq:G}
\end{align}
where the matrix elements of $\mathbb{A}_{\mu}$ are $\big(\mathbb{A}_{\mu}\big)^{\l}_{\n} = \Upsilon^{\l}_{\m\n}$.

The analog of the Riemann curvature tensor for the connection $\hat{\nabla}$ is 
\begin{align}
\mathcal{R}^{\k}_{\n\l\m} = \p_{\l} \Upsilon^{\k}_{\m\n} - \p_{\m} \Upsilon^{\k}_{\l\n} + \Upsilon^{\k}_{\l\a} \Upsilon^{\a}_{\m\n} - \Upsilon^{\k}_{\m\a} \Upsilon^{\a}_{\l\n} {.}
\end{align}
The covariant curvature tensor (field strength) $\mathcal{R}_{\k\n\l\m}=h_{\k\r} \mathcal{R}^{\r}_{\n\l\m}$ can be shown to be anti-Hermitian in the first pair of indices and anti-symmetric in the last pair.  This highlights a key difference with respect to the Riemann curvature tensor $R_{ijkl}$, which is real-valued and anti-symmetric in both the first and last pairs of indices.

\section{Adiabatic perturbation theory \label{sec:APT}}

We seek an approximate solution to the Schr\"odinger equation 
\begin{align}
i\hbar \p_t |\psi\rangle = H(x) |\psi\rangle
\label{eq:schroedinger:1}
\end{align}
of a quantum system with a time-dependent Hamiltonian of the form $H=H(x)$, where $x=x(t)$ is a path in parameter space that starts at $x_0$ at $t=0$ and ends at $x_1$ at $t=T$.  The larger $T$ is the more slowly the system is driven.  The lowest order approximation is 
\begin{align}
|\psi^{(0)}(t)\rangle = e^{-i \hbar^{-1} \int_0^t E_n(t') dt'} e^{i \int_0^t A_{\m}\dot{x}^{\m} dt'} \big|n\big(x(t)\big)\big> {,} 
\label{eq:soln:lowest}
\end{align}
where $|n(x)\rangle$ is the $n$th eigenstate of $H(x)$ with energy $E_n(x)$, assumed to be nondegenerate, $A_{\m} = i\langle n | \p_{\m} n\rangle$, and $\dot{x}^{\m}=dx^{\m}/dt$. 

To develop a systematic perturbation theory in the limit $T\rightarrow \infty$, we identify a dimensionless small parameter as follows.  Introducing a scaled time variable $s=t/T$ and a dimensionless Hamiltonian $h(x) = H(x)/\Delta$, where $\Delta$ is a characteristic energy scale of $H(x)$, and defining the dimensionless parameter
\begin{align}
\epsilon = \f{\hbar}{\Delta T} {,}
\end{align}
the Schr\"odinger equation becomes
\begin{align}
i\epsilon \p_s |\psi\rangle = h(x) |\psi\rangle {.}
\label{eq:schroedinger:2}
\end{align}
Since $\epsilon$ appears in this equation in exactly the same way that $\hbar$ appears in Eq.~(\ref{eq:schroedinger:1}), we return to Eq.~(\ref{eq:schroedinger:1}) and develop a perturbation series for the solution in powers of $\hbar$.  This perturbative solution is equivalent to the corresponding perturbation series for the solution of Eq.~(\ref{eq:schroedinger:2}) in powers of $\epsilon$.  We shall say that an approximate solution $|\psi^{(p)}\rangle$ is a $p$th order solution if it satisfies 
\begin{align}
\lim_{\hbar\rightarrow 0} \f{1}{\hbar^p} \big|\big| |\psi^{(p)}(t)\rangle - |\psi(t)\rangle \big|\big| = 0 \label{eq:condition}
\end{align}
for all times in the interval $[0,T]$.  According to this criterion, $|\psi^{(0)}\rangle$ is a 0th order solution. 

As an initial step toward the general $p$th order solution, we look for a first-order solution of the form
\begin{align}
|\psi^{(1)\prime}\rangle = e^{i\hbar^{-1}\phi} e^{i\gamma} \big( | n\rangle + \hbar| n_1^{\prime}\rangle \big) {,}
\label{eq:psi}
\end{align}
where $\phi=-\int^t E_n(t')dt'$ and $\gamma=\int^t A_{\m} \dot{x}^{\m} dt'$.  Substituting $|\psi^{(1)\prime}\rangle$ into the Schr\"odinger equation and collecting equal powers of $\hbar$ yields, at first order,
\begin{align}
(E_n-H)|n_1^{\prime}\rangle  &= -i |D_{\n} n\rangle \dot{x}^{\n} {.}
\end{align}
Since $|D_{\n} n\rangle$ has no $|n\rangle$ component, i.e.~$\langle n | D_{\n} n\rangle=0$, the resolvent $(E_n-H)^{-1}$ acts regularly on the gauge-covariant derivative of $|n\rangle$, and we have
\begin{align}
|n_1^{\prime}\rangle &= -i (E_n-H)^{-1} |T\rangle + (\beta_1 + i\alpha_1) |n\rangle {,}
\label{eq:n1prime}
\end{align}
where $|T\rangle = | D_{\n} n\rangle \dot{x}^{\n}$ is the tangent ket to the path $x=x(t)$.  The condition that $|\psi^{(1)}\rangle$ be normalized to first order in $\hbar$ implies $\beta_1=0$.  The coefficient $\alpha_1$ is determined by projecting the second-order terms in the Schr\"odinger equation onto $|n\rangle$, giving
\begin{align}
0 &= i \langle n |D_{\n} n_1^{\prime}\rangle \dot{x}^{\n} \nn \\
&= i\p_{t} \langle n | n_1^{\prime} \rangle - i \langle D_{\n} n |n_1^{\prime}\rangle \dot{x}^{\n} {,}
\end{align}
which implies
\begin{align}
\dot{\alpha}_1 &= \mathrm{Im} \langle T | n_1^{\prime} \rangle \nn \\
&= -\langle T | (E_n-H)^{-1} |T\rangle {.}
\label{eq:dalpha1}
\end{align}
Every term in Eq.~(\ref{eq:n1prime}) is gauge covariant and coordinate invariant.  Writing the first term as
\begin{align}
-i \sum_{m\neq n} \f{\langle m | \p_{\n} H | n\rangle \dot{x}^{\n}}{(E_n-E_m)^2} |m\rangle 
\end{align}
shows that it contains the nonadiabatic coupling. 

We continue to second order and look for an approximate solution of the form
\begin{align}
|\psi^{(2)\prime}\rangle = e^{i\hbar^{-1}\phi} e^{i\gamma} \big( | n\rangle + \hbar| n_1^{\prime}\rangle + \hbar^2 | n_2^{\prime}\rangle \big) {.}
\label{eq:psi}
\end{align}
Substitution into the Schr\"odinger equation and isolation of the second-order terms yields
\begin{align}
|n_2^{\prime}\rangle &= - (E_n-H)^{-2} \hat{\nabla}_{\bf T} | T\rangle \nn \\
&\quad - (E_n-H)^{-1} \p_t (E_n-H)^{-1} | T\rangle \nn \\
&\quad+ (\beta_2 + i\alpha_2) | n\rangle {,}
\label{eq:n2prime}
\end{align}
where
\begin{align}
\hat{\nabla}_{\bf T} | T\rangle = | D_{\n} n\rangle \ddot{x}^{\n} + | D_{\l} n\rangle \Upsilon_{\m\n}^{\l} \dot{x}^{\m} \dot{x}^{\n} 
\end{align}
is the quantum covariant derivative of $|T\rangle$ with respect to the tangent vector $\mathbf{T}$.
The first term on the right hand side of Eq.~(\ref{eq:n2prime}) is proportional to the purely intrinsic geometric quantity $\hat{\nabla}_{\bf T} | T\rangle$, which is fully determined by the local geometry induced by $|n(x)\rangle$ at every point along the path $x=x(t)$.  The $(E_n-H)^{-2}$ factor decreases the contribution of states with energy far from $E_n$.  The second term in Eq.~(\ref{eq:n2prime}) is a type of higher-order nonadiabatic coupling, depending quadratically on the velocities instead of linearly.  The coefficient $\beta_2$ in the third term ensures that $|\psi^{(2)}\rangle$ is normalized.
From the third-order terms in the Schr\"odinger equation, we find that $\alpha_2$ and $\beta_2$ must satisfy
\begin{align}
\dot{\beta}_2 + i \dot{\alpha}_2 &= - \langle n | D_{\n} n_2^{\prime} \rangle \dot{x}^{\n} \nn \\
&= \langle T | n_2^{\prime} \rangle {}
\label{eq:identity:dalpha:dbeta}
\end{align}
in order for $|\psi^{(2)\prime}\rangle$ to be a second-order solution according to Eq.~(\ref{eq:condition}).  Using Eq.~(\ref{eq:n2prime}), we have
\begin{align}
\dot{\beta}_2 + i \dot{\alpha}_2 &= - \langle T | (E_n-H)^{-2} \hat{\nabla}_{\bf T} | T\rangle \nn \\
&\quad - \langle T| (E_n-H)^{-1}\p_{t} (E_n-H)^{-1}| T\rangle {.}
\label{eq:dalpha2:dbeta2}
\end{align}
Instead of integrating $\int \dot{\beta}_2 dt$, it is simpler to obtain $\beta_2$ directly from the normalization condition for $|\psi^{(2)\prime}\rangle$, which implies
\begin{align}
\beta_2 &= -\f{1}{2} \langle T | (E_n-H)^{-2} | T \rangle {.}
\label{eq:beta2}
\end{align}

Before proceeding further, it is crucial to realize that since $i\alpha_1$ and $\beta_2+i\alpha_2$ are coefficients of $|n\rangle$, they can be absorbed into the overall exponential factor.  Indeed, the alternative second-order solution 
\begin{align}
|\psi^{(2)}\rangle &= e^{i\hbar^{-1}\phi} e^{i\gamma} e^{i\hbar\alpha_1}  e^{\hbar^2 (\beta_2+i\alpha_2)} \big(|n\rangle + \hbar |n_1\rangle + \hbar^2 |n_2\rangle \big) {}
\label{eq:psi2}
\end{align}
with 
\begin{align}
|n_1\rangle &= -i (E_n-H)^{-1} |T\rangle \nn \\
|n_2\rangle &= -(E_n-H)^{-2}\hat{\nabla}_{\bf T} |T\rangle \nn \\
&\quad - (E_n-H)^{-1} \p_t(E_n-H)^{-1} |T\rangle {}
\end{align}
is equivalent to $|\psi^{(2)\prime}\rangle$ through second order in $\hbar$.

We are now in a position to write the general $p$th-order solution
\begin{align}
| \psi^{(p)}\rangle = e^{i\hbar^{-1} \phi} e^{i\gamma} e^{\hbar(\beta_1+i\alpha_1)} \cdots \cdots e^{\hbar^p(\beta_p+i\alpha_p)}& \nn \\
\cdot \big( | n\rangle + \hbar | n_1\rangle + \cdots + \hbar^p | n_p\rangle \big) &
{.} \label{eq:solution:p}
\end{align}
The rationale behind Eq.~(\ref{eq:solution:p}) is to express the wave function as a product of three factors: (i) the factor $(| n\rangle + \hbar | n_1\rangle + \cdots + \hbar^p | n_p\rangle)$, which is a function of the instantaneous position along the path $x=x(t)$, (ii) an integrable exponential factor $e^{i\hbar^{-1}\phi} e^{\hbar \beta_1} \cdots e^{\hbar^p \beta_p}$ and (iii) a nonintegrable phase factor $e^{i\gamma} e^{i\hbar \alpha_1} \cdots e^{i\hbar^p \alpha_p}$, which is a path-dependent quantity.  The form of solution in Eq.~(\ref{eq:solution:p}) is convenient when evaluating the expectation value $\langle \psi^{(p)} | \hat{A} | \psi^{(p)}\rangle$ of an observable $\hat{A}$ because the nonintegrable phase factors cancel out: no time integrals need to be evaluated.

By substituting $|\psi^{(p)}\rangle$ into the Schr\"odinger equation, we can derive a recurrence relation that determines $|n_p\rangle$ in terms of the $|n_k\rangle$ with $k<p$.  
From the $\mathcal{O}(\hbar^p)$ terms, we find
\begin{align}
(E_n-H) |n_p\rangle &= - i |D_{\n} n_{p-1}\rangle \dot{x}^{\n} - i(\dot{\beta}_{p-1}+i\dot{\alpha}_{p-1}) |n\rangle \nn \\
&\quad - \sum_{k=1}^{p-2} i(\dot{\beta}_k+i \dot{\alpha}_k) |n_{p-1-k}\rangle {.}
\label{eq:intermediate}
\end{align}
Since Eq.~(\ref{eq:identity:dalpha:dbeta}) generalizes to
\begin{align}
\dot{\beta}_k + i \dot{\alpha}_k &= - \langle n | D_{\n} n_k \rangle \dot{x}^{\n} \nn\\
&= \langle T | n_k \rangle {,}
\end{align}
the second term on the right hand side of Eq.~(\ref{eq:intermediate}) removes the $|n\rangle$ component from the first term, i.e.~these two terms together equal  
\begin{align}
- i \big(1-|n\rangle\langle n|\big) |D_{\n} n_{p-1}\rangle \dot{x}^{\n} {.} 
\end{align}
Further, since the $|n\rangle$ component of $|n_k\rangle$ has been absorbed into the overall exponential factor at every order $k$, we have $\langle n|n_k\rangle = 0$ for $1\leq k \leq p$.  Therefore, the resolvent acts regularly on the right hand side of Eq.~(\ref{eq:intermediate}), and we obtain the recurrence relation
\begin{align}
|n_p\rangle &= - i (E_n-H)^{-1}\hat{\nabla}_{\bf T} |n_{p-1} \rangle \nn \\
&\quad -i \sum_{k=1}^{p-2} (\dot{\beta}_k+i \dot{\alpha}_k) (E_n-H)^{-1} |n_{p-1-k}\rangle {.}
\label{eq:recurrence}
\end{align}
Equation (\ref{eq:recurrence}) is expressed in a manifestly geometric form owing to the use of the quantum covariant derivative $\hat{\nabla}_{\bf T}$.  It is convenient to record here the third-order correction
\begin{align}
|n_3\rangle &= i(E_n-H)^{-3}\hat{\nabla}_{\bf T}\hat{\nabla}_{\bf T} |T\rangle \nn \\
&\quad + i (E_n-H)^{-1} \p_t (E_n-H)^{-2}\hat{\nabla}_{\bf T} |T\rangle \nn \\
&\quad + i (E_n-H)^{-2} \p_t (E_n-H)^{-1}\hat{\nabla}_{\bf T} |T\rangle \nn \\
&\quad + i (E_n-H)^{-1}\p_t \big[ (E_n-H)^{-1} \p_t (E_n-H)^{-1} \big] |T\rangle \nn\\
&\quad + \dot{\alpha}_1 (E_n-H)^{-1} |n_1\rangle {}
\end{align}
and the third-order normalization coefficient
\begin{align}
\beta_3 &= - \mathrm{Re} \langle n_1 | n_2 \rangle \nn \\
&= -\mathrm{Im} \langle T | (E_n-H)^{-3}\hat{\nabla}_{\bf T} |T\rangle \nn \\
&\quad - \mathrm{Im} \langle T | (E_n-H)^{-2} \p_t (E_n-H)^{-1} |T\rangle {,}
\end{align}
which have not been derived previously.
The expectation value of the Hamiltonian, to third order, is
\begin{align}
E_n^{(3)} &= E_n + \f{1}{2} \mathcal{M}_{2\m\n} \dot{x}^{\mu} \dot{x}^{\nu} + E_{n3} {,}
\end{align}
where
\begin{align}
\mathcal{M}_{2\mu\nu} = 2 \hbar^2 \mathrm{Re} \langle D_{\m} n |  (H-E_n)^{-1} |  D_{\n} n\rangle
\end{align}
is an induced inertia (mass) tensor, which has appeared previously \cite{inglis1954,littlejohn1993,goldhaber2005,requist2010,scherrer2017}.  The third-order perturbation is
\begin{align}
E_{n3}
&= -2\hbar^3 \mathrm{Im} \langle T | (E_n-H)^{-2}\hat{\nabla}_{\bf T} | T\rangle \nn \\
&\quad -2\hbar^3  \mathrm{Im} \langle T | (E_n-H)^{-1} \p_t (E_n-H)^{-1} | T\rangle {.}
\label{eq:En3}
\end{align}
If the quantum system described by $|\psi(x)\rangle$ is coupled to a ``heavy'' classical system responsible for the slow motion of $x$, the third-order corrections generate non-Lagrangian corrections to the effective classical equation of motion. 

\section{Conclusions}

We have introduced a gauge- and coordinate-invariant adiabatic perturbation theory.  Since all nonintegrable phases appear in a single overall prefactor, this theory is very convenient for evaluating observables to high order because it is not necessary to evaluate any time integrals.  Just as first-order adiabatic perturbation theory was used to derive topological formulations of the quantum Hall conductance \cite{thouless1982} and adiabatic charge pumping \cite{thouless1983,niu1984} and a Berry phase formula for the macroscopic polarization \cite{king-smith1993}, the higher-order adiabatic perturbation theory developed here can be used to derive convenient formulas for nonlinear response properties.  

In summary, the quantum covariant derivative $\hat{\nabla}$ encodes a geometric structure beyond that described by the quantum geo\-metric tensor.  The quantum covariant derivative is precisely the geometric structure needed to parallel transport wave function tangent vectors, and it is indispensable in the derivation of adiabatic perturbation theory. 

\appendix

\section{Parallel transport in a three-state system}\label{secA1}

Consider the generic state of a three-state system parametrized up to an overall phase as 
\begin{align}
|\psi\rangle = \left( \begin{array}{l} \sin\theta \sin\beta e^{-i\gamma} e^{i\alpha} \\[0.1cm] \sin\theta \cos\beta e^{-i\gamma} e^{-i\alpha} \\[0.1cm] \cos\theta \end{array} \right) {,} \label{eq:ground:state}
\end{align}
where the angular variables $\alpha,\beta,\gamma,\theta$ depend parametrically on a two-dimensional configuration space $M$, i.e.~the coordinates $(x^1,x^2)$ are local coordinates for the manifold $M$.
This state can be obtained by acting with a sequence of unitary rotations on a reference state as follows
\begin{align}
|\psi\rangle = e^{i \lambda_3 \alpha} e^{i \lambda_2 \beta} e^{i \lambda_3 \gamma} e^{i \lambda_7 \theta} \left( \begin{array}{c} 0 \\ 0 \\ 1 \end{array} \right) {,}
\end{align}
where $\lambda_i$ are the Gell-Mann matrices and $\alpha,\beta,\gamma,\theta$ are four out of a set of eight generalized Euler angles that parametrize the $SU(3)$ Lie group \cite{byrd1998,byrd2000}.

From the gauge connection one-form 
\begin{align}
A &= i \langle \psi | d\psi \rangle \nn \\
&= \sin^2\theta d\gamma +\sin^2\theta \cos2\beta d\alpha {,}
\end{align}
we read off the following canonically conjugate coordinates:
\begin{align}
q^1 &= \gamma  & p_1 &= \sin^2\theta \nn \\
q^2 &= \alpha & p_2 &= \sin^2\theta \cos2\beta {.}
\end{align}
We will use the notation $\xi^{\mu}$ for the tuple of variables $\{q^1,q^2,p_1,p_2\}$.  In terms of these coordinates, the state is 
\begin{align}
|\psi\rangle = \left( \begin{array}{l} \sqrt{\f{p_1-p_2}{2}} e^{-iq_1} e^{iq_2} \\[0.1cm] \sqrt{\f{p_1+p_2}{2}} e^{-iq_1} e^{-iq_2} \\[0.1cm] \sqrt{1-p_1} \end{array} \right) {.} \label{eq:ground:state:qp}
\end{align}
The canonical coordinates define a gauge-covariant coordinate frame $\{| D_m\psi\rangle\}$ with
\begin{align}
| D_{m}\psi\rangle &= |\p_{m}\psi\rangle - |\psi\rangle \langle \psi | \p_{m} \psi\rangle {.}
\end{align}
We find
\begin{align}
{\scriptstyle | D_1\psi\rangle } &= \left( \begin{array}{c} {\scriptstyle -i (1-p_1) \sqrt{(p_1-p_2)/2} e^{-i (q_1-q_2)} } \\ {\scriptstyle -i (1-p_1) \sqrt{(p_1+p_2)/2} e^{-i (q_1+q_2)} } \\ {\scriptstyle i p_1 \sqrt{1-p_1} } \end{array} \right) \nn \\
{\scriptstyle | D_2\psi\rangle } &= \left( \begin{array}{c} {\scriptstyle i (1+p_2) \sqrt{(p_1-p_2)/2} e^{-i (q_1-q_2)} } \\ {\scriptstyle -i (1-p_2) \sqrt{(p_1+p_2)/2} e^{-i (q_1+q_2)} } \\ {\scriptstyle i p_2 \sqrt{1-p_1} } \end{array} \right) \nn \\
{\scriptstyle | D_3\psi\rangle } &= \left( \begin{array}{c} {\scriptstyle (1/4) \sqrt{2/(p_1-p_2)} e^{-i (q_1-q_2)} } \\ {\scriptstyle (1/4) \sqrt{2/(p_1+p_2)} e^{-i (q_1+q_2)} } \\ {\scriptstyle -(1/2) \sqrt{1/(1-p_1)} } \end{array} \right) \nn \\ 
{\scriptstyle | D_4\psi\rangle } &= \left( \begin{array}{c} {\scriptstyle -(1/4) \sqrt{2/(p_1-p_2)} e^{-i (q_1-q_2)} } \\ {\scriptstyle (1/4) \sqrt{2/(p_1+p_2)} e^{-i (q_1+q_2)} } \\ {\scriptstyle 0 } \end{array} \right) {.}
\end{align}
The quantum metric $g_{mn} = \mathrm{Re}\langle D_{m} \psi | D_{n} \psi \rangle$ in canonical coordinates is
\begin{align}
{\scriptstyle (g_{mn}) = \left( \begin{array}{cccc} 
{\scriptstyle p_1 (1-p_1) } & {\scriptstyle p_2 (1-p_1) } & {\scriptstyle 0 } & {\scriptstyle 0 } \\ 
{\scriptstyle p_2 (1-p_1) } & {\scriptstyle p_1 - p_2^2 } & {\scriptstyle 0 } & {\scriptstyle 0 } \\
{\scriptstyle 0 } & {\scriptstyle 0 } & {\scriptstyle \f{p_1-p_2^2}{4(1-p_1)(p_1^2-p_2^2)} } & {\scriptstyle -\f{p_2}{4(p_1^2-p_2^2)} } \\
{\scriptstyle 0 } & {\scriptstyle 0 } & {\scriptstyle -\f{p_2}{4(p_1^2-p_2^2)} } & {\scriptstyle \f{p_1}{4(p_1^2-p_2^2)} } 
\end{array} \right) }
\label{eq:g:matrix}
\end{align}
and the Berry curvature $B_{mn} = -2\mathrm{Im}\langle D_{m} \psi | D_{n} \psi \rangle$ is
\begin{align}
(B_{mn}) = \left( \begin{array}{cccc} 
0 & 0 & -1 & 0 \\ 
0 & 0 & 0 & -1 \\
1 & 0 & 0 & 0 \\
0 & 1 & 0 & 0
\end{array} \right) {.}
\label{eq:B:matrix}
\end{align}
It is instructive to compare Eq.~(\ref{eq:g:matrix}) with the result for a two-level system in the state $(\cos(\theta/2) , \sin(\theta/2) e^{i\varphi})^T$:
\begin{align}
(g_{mn}) = \left( \begin{array}{cc} p(1-p) & 0 \\ 0 & \f{1}{4p(1-p)} \end{array} \right) {,}
\label{eq:g:twolevel}
\end{align}
in the $(q,p)$ basis, where $q=\varphi$, $p=\sin^2(\theta/2)$, and $\theta,\varphi$ are the usual Bloch sphere angles.  Poles and zeros of the elements of $g$ occur for the values $p=1$ and $p=0$; these are singular points of the coordinate chart where either the first or second element of the state vanishes.  We also observe poles and zeros of the elements of $g$ in Eq.~(\ref{eq:g:matrix}) corresponding to the vanishing of elements of the three-level state in Eq.~(\ref{eq:ground:state:qp}). It is readily verified that after choosing $p_1=1$, which sets the third element of $|\psi\rangle$ to zero, and transforming to the variables $q=2q_2$ and $p=(1+p_2)/2$, the relevant block of the three-level $g$ in Eq.~(\ref{eq:g:matrix}) simplifies to Eq.~(\ref{eq:g:twolevel}).  In terms of our choice of canonical coordinates, $h_{mn}= g_{mn} - \f{i}{2} B_{mn}$ only depends on the $p_{m}$ variables and not the $q^{m}$ variables, which is useful for calculations.  The Berry curvature two-form $B=B_{mn} d\xi^{m} \wedge d\xi^{n}$ is related to the canonical symplectic two-form $\omega=dp_{m} \wedge dq^{m}$ according to $\omega = -\f{1}{2} B$.

Now we turn to the evaluation of geometric quantities associated with the manifold $M$.  For tensors, this is conveniently done by pulling back the tensor in canonical $q,p$ coordinates to the tensor in terms of the $(x^1,x^2)$ coordinates on the configuration manifold $M$ \cite{requist2016b}.  The quantum metric in $(x^1,x^2)$ coordinates is
\begin{align*}
g_{\mu\nu} &= p_1 (1-p_1) q_{1\m} q_{1\n} + p_2 (1-p_1) (q_{1\m} q_{2\n} + q_{2\m} q_{1\n}) \nn \\
&+ (p_1-p_2^2) q_{2\m} q_{2\n} + \f{p_1-p_2^2}{4(1-p_1)(p_1^2-p_2^2)} p_{1\m} p_{1\n} \nn \\
&- \f{p_2}{4(p_1^2-p_2^2)} (p_{1\m} p_{2\n} + p_{2\m} p_{1\n}) + \f{p_1}{4(p_1^2-p_2^2)} p_{2\m} p_{2\n} 
\end{align*}
and the Berry curvature is
\begin{align}
B_{\mu\nu} &= p_{1\m} q_{1\n} - p_{1\n} q_{1\m} + p_{2\m} q_{2\n} - p_{2\n} q_{2\m} \nn \\
&= \{p_1,q_1\} + \{p_2,q_2\} {,}
\end{align}
where e.g.~$q_{1\m}=\p_{\m} q_1$ and $\{q_1,p_1\}$ is the Poisson bracket with respect to $x^\m$ variables.
We will need the inverse of $(h_{\mu\nu})$, which is
\begin{align}
(h^{\mu\nu}) = \f{1}{\det(h_{\mu\nu})} \left( \begin{array}{cc} h_{22} & -h_{12} \\ -h_{21} & h_{11} \end{array} \right)
\label{eq:h:inverse}
\end{align}
with the determinant 
\begin{align*}
{\scriptstyle
\det(h_{\mu\nu}) } &= 
{\scriptstyle \f{1}{16(1-p_1)(p_1^2-p_2^2)} \Big\{ } \nn\\
&\quad {\scriptstyle 8 p_{11} p_{22} (1-p_1) (p_1 q_{12} + p_2 q_{22}) [p_2(1-p_1) q_{11} + (p_1-p_2^2) q_{21}] } \nn \\
&{\scriptstyle + 8 p_{12} p_{21} (1-p_1) (p_1 q_{11} + p_2 q_{21}) [p_2(1-p_1) q_{12} + (p_1-p_2^2) q_{22}] } \nn \\
&{\scriptstyle - 8 p_{11} p_{21} (1-p_1) (p_1 q_{12} + p_2 q_{22}) [p_2(1-p_1) q_{12} + (p_1-p_2^2) q_{22}] } \nn \\
&{\scriptstyle - 8 p_{12} p_{22} (1-p_1) (p_1 q_{11} + p_2 q_{21}) [p_2(1-p_1) q_{11} + (p_1-p_2^2) q_{21}] } \nn \\
&{\scriptstyle - 8 p_{21} p_{22} (1-p_1)^2 (p_1 q_{11} + p_2 q_{21}) (p_1 q_{12} + p_2 q_{22}) } \nn \\
&{\scriptstyle - 8 p_{11} p_{12} [p_2(1-p_1) q_{11} + (p_1-p_2^2) q_{21}] [p_2(1-p_1) q_{12} + (p_1-p_2^2) q_{22}] } \nn \\
&{\scriptstyle + 4 p_{11}^2 [p_2(1-p_1) q_{12} + (p_1-p_2^2) q_{22}]^2 } \nn \\
&{\scriptstyle + 4 p_{12}^2 [p_2(1-p_1) q_{11} + (p_1-p_2^2) q_{21}]^2 } \nn \\
&{\scriptstyle + 4 p_{21}^2 (1-p_1)^2 (p_1 q_{12} + p_2 q_{22})^2 } \nn \\
&{\scriptstyle + 4 p_{22}^2 (1-p_1)^2 (p_1 q_{11} + p_2 q_{21})^2 } \nn \\
&{\scriptstyle + p_{11}^2 p_{22}^2 + p_{12}^2 p_{21}^2 - 2 p_{11} p_{12} p_{21} p_{22} \Big\} } \nn \\
&{\scriptstyle +(1-p_1) (p_1^2-p_2^2) (q_{11} q_{22} - q_{12} q_{21})^2 {.} } 
\end{align*}
Now we evaluate $\Upsilon_{\lambda\mu\nu} = \Gamma_{\lambda\mu\nu} + i C_{\lambda\mu\nu}$. The Christoffel symbol of the first kind of Riemannian geometry is readily evaluated with the knowledge of $g_{\m\n}$ to give
\begin{align*}
{\scriptstyle \Gamma_{\lambda\mu\nu} } &= 
{\scriptstyle \f{1-2p_1}{2} q_{1\l} (q_{1\m} p_{1\n}+p_{1\m} q_{1\n}) - \f{p_2}{2} q_{1\l} (q_{2\m} p_{1\n}+p_{1\m} q_{2\n}) } \nn \\
&{\scriptstyle + \f{1-p_1}{2} q_{1\l} (q_{2\m} p_{2\n}+p_{2\m} q_{2\n}) -\f{p_2}{2} q_{2\l} (q_{1\m} p_{1\n}+p_{1\m} q_{1\n}) } \nn \\
&{\scriptstyle + \f{1-p_1}{2} q_{2\l} (q_{1\m} p_{2\n}+p_{2\m} q_{1\n}) + \f{1}{2} q_{2\l} (q_{2\m} p_{1\n}+p_{1\m} q_{2\n}) } \nn \\
&{\scriptstyle - p_2 q_{2\l} (q_{2\m} p_{2\n}+p_{2\m} q_{2\n}) -\f{1-2p_1}{2} p_{1\l} q_{1\m} q_{1\n} } \nn \\
&{\scriptstyle + \f{p_2}{2} p_{1\l} (q_{1\m} q_{2\n}+q_{2\m} q_{1\n}) - \f{1}{2} p_{1\l} q_{2\m} q_{2\n} } \nn \\
&{\scriptstyle + \left( \f{1}{8(1-p_1)^2} -\f{p_1^2+p_2^2}{8(p_1^2-p_2^2)^2} \right) p_{1\l} p_{1\m} p_{1\n} } \nn \\
&{\scriptstyle + \f{p_1 p_2}{4(p_1^2-p_2^2)^2} p_{1\l} (p_{1\m} p_{2\n}+p_{2\m} p_{1\n}) - \f{p_1^2+p_2^2}{8(p_1^2-p_2^2)^2} p_{1\l} p_{2\m} p_{2\n} } \nn \\
&{\scriptstyle -\f{1-p_1}{2} p_{2\l} (q_{1\m} q_{2\n}+q_{2\m} q_{1\n}) + p_2 p_{2\l} q_{2\m} q_{2\n} } \nn \\
&{\scriptstyle + \f{p_1 p_2}{4(p_1^2-p_2^2)^2} p_{2\l} p_{1\m} p_{1\n} - \f{p_1^2+p_2^2}{8(p_1^2-p_2^2)^2} p_{2\l} (p_{1\m} p_{2\n}+p_{2\m} p_{1\n}) } \nn \\
&{\scriptstyle + \f{p_1 p_2}{4(p_1^2-p_2^2)^2} p_{2\l} p_{2\m} p_{2\n} +\f{p_1-p_2^2}{4(1-p_1)(p_1^2-p_2^2)} p_{1\l} p_{1\m\n} } \nn \\
&{\scriptstyle - \f{p_2}{4(p_1^2-p_2^2)} (p_{1\l} p_{2\m\n} + p_{2\l} p_{1\m\n}) + \f{p_1}{4(p_1^2-p_2^2)} p_{2\l} p_{2\m\n} } \nn \\
&{\scriptstyle +p_1(1-p_1) q_{1\l} q_{1\m\n} + p_2 (1-p_1) (q_{1\l} q_{2\m\n} + q_{2\l} q_{1\m\n}) } \nn \\
&{\scriptstyle + (p_1-p_2^2) q_{2\l} q_{2\m\n} {,} }
\end{align*}
where e.g.~$p_{1\m\n}=\p_{\n}\p_{\m}p_1$. 
The symmetry of $\Gamma_{\lambda\mu\nu}$ with respect to interchange of the second and third indices is transparent.

To evaluate $C_{\lambda\mu\nu}$, we use the identity \cite{requist2022a}
\begin{align}
C_{\lambda\mu\nu} &= \mathrm{Im} \langle \p_{\lambda}\psi | \p_{\mu} \p_{\nu} \psi \rangle + A_{\lambda} g_{\mu\nu} + A_{\mu} g_{\lambda\nu} + A_{\nu} g_{\lambda\mu} \nn \\
&\quad+ A_{\lambda} A_{\mu} A_{\nu} {,}
\end{align}
and find
\begin{align}
{\scriptstyle C_{\l\m\n} } &=
{\scriptstyle q_{1\l} q_{1\m} [ p_1 (1-p_1) (1-2p_1) q_{1\n} + p_2 (1-p_1) (1-2p_1) q_{2\n} ] } \nn \\
&{\scriptstyle +q_{1\l} q_{2\m} [ p_2 (1-p_1) (1-2p_1) q_{1\n} + (1-p_1) (p_1-2p_2^2) q_{2\n} ] } \nn \\
&{\scriptstyle +q_{1\l} p_{1\m} \Big[ \f{p_1(1-2p_1+p_2^2)}{4(1-p_1)(p_1^2-p_2^2)} p_{1\n} - \f{p_2 (1-p_1)}{4(p_1^2-p_2^2)} p_{2\n} \Big] } \nn \\
&{\scriptstyle +q_{1\l} p_{2\m} \Big[ -\f{p_2 (1-p_1)}{4(p_1^2-p_2^2)} p_{1\n} + \f{p_1 (1-p_1)}{4(p_1^2-p_2^2)} p_{2\n} \Big] } \nn \\
&{\scriptstyle +q_{2\l} q_{1\m} [ p_2 (1-p_1)(1-2p_1) q_{1\n} + (1-p_1)(p_1-2p_2^2) q_{2\n} ] } \nn \\
&{\scriptstyle +q_{2\l} q_{2\m} [ (1-p_1)(p_1-2p_2^2)q_{1\n} + p_2 (1-3p_1+2p_2^2) q_{2\n} ] } \nn \\
&{\scriptstyle +q_{2\l} p_{1\m} \Big[ -\f{p_2(1-p_2^2)}{4(1-p_1)(p_1^2-p_2^2)} p_{1\n} + \f{p_1+p_2^2}{4(p_1^2-p_2^2)} p_{2\n} \Big] } \nn \\
&{\scriptstyle +q_{2\l} p_{2\m} \Big[ \f{p_1+p_2^2}{4(p_1^2-p_2^2)} p_{1\n} - \f{p_2(1+p_1)}{4(p_1^2-p_2^2)} p_{2\n} \Big] } \nn \\
&{\scriptstyle +p_{1\l} q_{1\m} \Big[ \f{p_1(1-2p_1+p_2^2)}{4(1-p_1)(p_1^2-p_2^2)} p_{1\n} - \f{p_2 (1-p_1)}{4(p_1^2-p_2^2)} p_{2\n} \Big] } \nn \\
&{\scriptstyle +p_{1\l} q_{2\m} \Big[ -\f{p_2 (1-p_2^2)}{4(1-p_1)(p_1^2-p_2^2)} p_{1\n} + \f{p_1+p_2^2}{4(p_1^2-p_2^2)} p_{2\n} \Big] } \nn \\
&{\scriptstyle +p_{1\l} p_{1\m} \Big[ \f{p_1(1-2p_1+p_2^2)}{4(1-p_1)(p_1^2-p_2^2)} q_{1\n} - \f{p_2(1-p_2^2)}{4(1-p_1)(p_1^2-p_2^2)} q_{2\n} \Big] } \nn \\
&{\scriptstyle +p_{1\l} p_{2\m} \Big[ -\f{p_2(1-p_1)}{4(p_1^2-p_2^2)} q_{1\n} + \f{p_1+p_2^2}{4(p_1^2-p_2^2)} q_{2\n} \Big] } \nn \\
&{\scriptstyle +p_{2\l} q_{1\m} \Big[ -\f{p_2(1-p_1)}{4(p_1^2-p_2^2)} p_{1\n} + \f{p_1(1-p_1)}{4(p_1^2-p_2^2)} p_{2\n} \Big] } \nn \\
&{\scriptstyle +p_{2\l} q_{2\m} \Big[ \f{p_1+p_2^2}{4(p_1^2-p_2^2)} p_{1\n} - \f{p_2 (1+p_1)}{4(p_1^2-p_2^2)} p_{2\n} \Big] } \nn \\
&{\scriptstyle +p_{2\l} p_{1\m} \Big[ -\f{p_2 (1-p_1)}{4(p_1^2-p_2^2)} q_{1\n} + \f{p_1+p_2^2}{4(p_1^2-p_2^2)} q_{2\n} \Big] } \nn \\
&{\scriptstyle +p_{2\l} p_{2\m} \Big[ \f{p_1 (1-p_1)}{4(p_1^2-p_2^2)} q_{1\n} - \f{p_2 (1+p_1)}{4(p_1^2-p_2^2)} q_{2\n} \Big] } \nn \\
&{\scriptstyle +\f{1}{2} (p_{1\l} q_{1\m\n} - q_{1\l} p_{1\m\n} + p_{2\l} q_{2\m\n} - q_{2\l} p_{2\m\n}) {.} }
\end{align}

We have computed the quantum Christoffel symbol of the second kind using the formula
\begin{align}
\Upsilon^{\lambda}_{\mu\nu} = h^{\lambda\kappa} \Upsilon_{\kappa\mu\nu} {,}
\label{eq:Upsilon:second}
\end{align}
the Riemann curvature tensor
\begin{align}
\mathcal{R}^{\k}_{\n\l\m} 
= \p_{\l} \Upsilon^{\k}_{\m\n} - \p_{\m} \Upsilon^{\k}_{\l\n} + \Upsilon^{\k}_{\l\r} \Upsilon^{\r}_{\m\n} - \Upsilon^{\k}_{\m\r} \Upsilon^{\r}_{\l\n} {,}
\end{align}
and the fully covariant Riemann curvature tensor
\begin{align}
\mathcal{R}_{\k\n\l\m} = h_{\k\r} \mathcal{R}^{\r}_{\n\l\m} {.}
\end{align}
The formulas are too lengthy to record here, but we have verified that $\mathcal{R}_{\k\n\l\m}$ is anti-Hermitian in its first pair of indices ($\k\n$) and anti-symmetric in its second pair of indices ($\l\m$).  The symbol $\Upsilon^{\lambda}_{\mu\nu}$, determined by Eqs.~(\ref{eq:h:inverse}) and (\ref{eq:Upsilon:second}), completely specifies the parallel transport law for a tangent ket $| u\rangle = u^{\l} | D_{\l} n\rangle$ along a path $x=x(s)$:
\begin{align}
\f{du^{\l}}{ds} + \f{dx^{\m}}{ds} \Upsilon^{\l}_{\m\n} u^{\n} = 0 {.}
\end{align}

\bibliography{coor-inv}

\end{document}